\documentclass{article}
\usepackage{spconf,amsmath,epsfig}
\usepackage{multirow}
\usepackage{booktabs, arydshln}
\usepackage{comment}
\usepackage{caption}

\let\OLDthebibliography\thebibliography
\renewcommand\thebibliography[1]{
  \OLDthebibliography{#1}
  \setlength{\parskip}{0pt}
  \setlength{\itemsep}{0pt plus 0.3ex}
}

\begin{document}\sloppy

\def\x{{\mathbf x}}
\def\L{{\cal L}}

\title{COMPLEXITY-BOOSTED ADAPTIVE TRAINING FOR BETTER LOW-RESOURCE ASR PERFORMANCE}
%
\name{Hongxuan Lu\textsuperscript{1} \thanks{First author and second author contribute equally to this work.}, Shenjian Wang\textsuperscript{2}, Biao Li\textsuperscript{2}}

\address{Beijing OPPO Telecommunications Corp., Ltd., Beijing, China}


\ninept

\maketitle

\begin{abstract}
During the entire training process of the ASR model, the intensity of data augmentation and the approach of calculating training loss are applied in a regulated manner based on preset parameters. For example, SpecAugment employs a predefined strength of augmentation to mask parts of the time-frequency domain spectrum. Similarly, in CTC-based multi-layer models, the loss is generally determined based on the output of the encoder's final layer during the training process. However, ignoring dynamic characteristics may suboptimally train models. To address the issue, we present a two-stage training method, known as complexity-boosted adaptive (CBA) training. It involves making dynamic adjustments to data augmentation strategies and CTC loss propagation based on the complexity of the training samples. In the first stage, we train the model with intermediate-CTC-based regularization and data augmentation without any adaptive policy. In the second stage, we propose a novel adaptive policy, called MinMax-IBF, which calculates the complexity of samples. We combine the MinMax-IBF policy to data augmentation and intermediate CTC loss regularization to continue training. The proposed CBA training approach shows considerable improvements, up to 13.4\% and 14.1\% relative reduction in WER on the LibriSpeech 100h test-clean and test-other dataset and also up to 6.3\% relative reduction on AISHELL-1 test set, over the Conformer architecture in Wenet.
\end{abstract}
\begin{keywords}
ASR, Data Augmentation, Intermediate CTC, WeNet, Conformer, Adaptive Training
\end{keywords}
\section{Introduction}

Automatic Speech Recognition (ASR) has witnessed remarkable advancements in recent years by well-designed self-attention-based End-to-end models. Besides the model architecture, the performance of ASR systems relies heavily on the availability of large and diverse training data with transcripts. The performance often comes with great computational cost \cite{chan2016listen} \cite{chorowski2015attention}. Considerable research efforts have been dedicated to enhancing ASR performance through advancements in data augmentation techniques \cite{bartelds2023making,casanova2022asr,lam2021fly, hu2021sapaugment, park2019specaugment} and model architecture \cite{kim2022squeezeformer, kim2023branchformer, ren2022improving, burchi2021efficient}.  

Regarding data augmentation (DA), multiple methods have been proposed and proved effective in the past, such as speed perturbation \cite{ko2015audio}, noise insertion \cite{6639100},  and reverberation \cite{7953152}. SpecAugment \cite{park2019specaugment} is one of the state-of-the-art approaches, which uses hand-crafted policies including warping features, and masking on time and frequency domain, while, the SpecAugment method is inflexible. SapAugment \cite{hu2021sapaugment} takes this idea a step further. It adapts the strength of augmentation automatically depending on the loss of samples and effectively combines different augmentation strategies.

\begin{figure}[h]
\centering
\includegraphics[width=8cm]{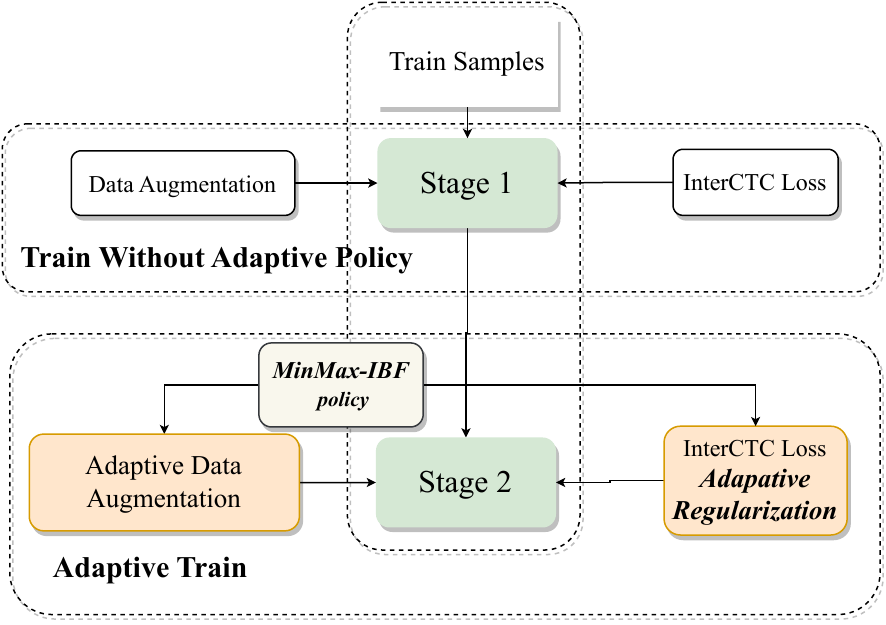}
\caption{The two stages of proposed CBA training.}
\label{fig:label}
\end{figure}

In terms of model architectures, various types of neural networks have been explored in recent years: Connectionist Temporal Classification (CTC) \cite{graves2006connectionist}, recurrent neural network transducer (RNN-T) \cite{graves2012sequence} and attention-based encoder-decoder (AED) \cite{vaswani2017attention}. Various auxiliary loss functions have been explored for different architectures \cite{toshniwal2017multitask} \cite{lee2021intermediate}. However, the model's training strategy remains rigid regardless of the complexity of training samples. 


In this paper, we adopt a two-stage training approach, as shown in Figure 1. Firstly, we propose the MinMax-IBF policy that calculates the complexity of samples, leading to a more accurate amount of augmentation. Instead of ranking the loss in a batch, our normalization policy distinguishes between complex and simple samples well. This gives better instructions on augmentation methods. Secondly, we introduce a loss fusion strategy, which combines the original loss and the proposed auxiliary loss. The auxiliary loss scalar weight is decided by the complexity of training samples. 
In this way, the model is well-regularized during the training process. Our experiments are based on WeNet Conformer \cite{gulati2020conformer} \cite{yao2021wenet}. The contributions of this paper are as follows:


\begin{itemize}
    \item For the first time, we propose to perform sample-adaptive training simultaneously on data augmentation and internal model modules.
    \item We propose a two-stage complexity-boosted adaptive training methodology where sample loss jointly decides the intensity of data augmentation and intermediate CTC regularization.
    \item The proposed method outperforms WeNet Conformer with SpecAugment by relatively 13.4\% (test-clean set) and 14.1\% (test-other set) on LibriSpeech. Furthermore, the proposed method achieves a relative 6.3\% CER reduction on the AISHELL-1 dataset.
\end{itemize}


\section{Related Work}

\noindent\textbf{SpecAugment.} It introduces feature domain time masking and frequency masking to data augmentation. It simply masks out segments of sizes \(\mathbf{m}_{t}\), and \(\mathbf{m_f}\), along the time and the frequency axes, respectively. SpecAugment is applied to Listen, Attend, and Spell networks for end-to-end speech recognition tasks, surpassing all previous studies in performance.

\noindent\textbf{SapAugment.} Previous observations have indicated that during training, the training samples are subjected to the same combination of augmentation methods, regardless of the complexity of those samples. Intuitively, a challenging sample should undergo a subtle augmentation, while an easy sample should undergo a more pronounced augmentation. This approach ensures that suitable training signals are generated to update the model parameters. SapAugment provides a sample-adaptive policy that perturbs samples according to their loss and applies a policy-based augmentation to train the ASR system. The loss-based policy, \(\mathbf{f}\), is parameterized by two hyperparameters, \(\mathbf{s}\), \(\mathbf{a}\), and is deﬁned using an \(\mathbf{IBF}\) (incomplete beta function)  \cite{incompletBeta}:

\begin{equation}
    f(loss)= 1 - IBF(s, a; loss_{\text{rank}} / B)
\end{equation}

where \(\mathbf{loss}_{\text{rank}}\) = 1,2,...,\(\mathbf{B}\) is the ranking of the loss value \(\mathbf{l}\) in a mini-batch, \(\mathbf{B}\) is the mini-batch size.

SapAugment also introduces five augmentations for ASR including feature domain augmentations, each with its own formula to determine the extent of augmentation based on the policy value.

\begin{table}[h]\footnotesize
\vspace{-0.2cm}
\centering
\caption{Adaptive strategy visualization: the correlation between the sample loss value and the data augmentation tensity.}
\renewcommand{\arraystretch}{1.3}
\begin{tabular}{c | c | c | c}
\hline
Loss Rank & Difficulty & f(loss) & SpecAug Mask Num\\ 
\hline
$\uparrow$  & harder & $\downarrow$ & $\downarrow$ \\
$\downarrow$ & simpler & $\uparrow$ & $\uparrow$  \\
\hline
\end{tabular}
\vspace{-0.2cm}
\end{table}
In Table 1, a higher rank of sample loss within a batch signifies a comparatively difficult sample to learn, like fast speech or high noise. By reducing the number of mask used in SpecAugment, the model can acquire a more comprehensive understanding of the information in such samples, and vice versa.

\noindent\textbf{Intermediate CTC loss.} 
CTC calculates the likelihood of target sequence \(\mathbf{y}\) by taking into account all possible alignments for the label and the input length \(\mathbf{T}\). Given the encoder output \(\mathbf{x}\), the likelihood is defined as:
\begin{equation}
    P_{\text{CTC}}(y|x):=\sum_{a \in \mathcal{B}^{-1}(y)} P(a|x)
\end{equation}
where \(\mathcal{B}^{-1}(\mathbf{y})\) is the set of alignment \(\mathbf{a}\) of length \(\mathbf{T}\) compatible to \(\mathbf{y}\) including the blank token.
The training process minimizes the negative log-likelihood using \(\mathbf{P}_{\text{CTC}}\) above:
\begin{equation}
    L_{\text{CTC}}:= -log P_{\text{CTC}} (y|x)
\end{equation}

For a multi-layer CTC-based network, \cite{lee2021intermediate} choose a layer, named "intermediate layer", and induce a sub-model by skipping all the layers after this layer. Training this sub-model regularizes the lower part of the full model. The CTC loss for this sub-model is similar to the full model:
\begin{equation}
    L_{\text{InterCTC}}:= -log P_{\text{CTC}} (y|x_{\text{InterLayer}})
\end{equation}

The total training loss is the weighted sum of the sub-model and full model loss, where the weight \(\lambda\) is a hyper-parameter :
\begin{equation}
    L_{\text{total}} = (1-\lambda)L_{\text{CTC}} +  \lambda L_{\text{InterCTC}}
\end{equation}

\section{PROPOSED ADAPTIVE TRAINING METHOD}

SpecAugment employs a fixed strategy without adaptive adjustments based on the samples. The operation of the mask remains at twice in WeNet Conformer. SapAugment takes initial steps to tackle this issue. However, the policy based on loss rank fails to consider the variability of individual samples within a batch, as mentioned in Equation 1. Additionally, the adaptive policy is only implemented in the data augmentation, overlooking other aspects of the model training.

Being inspired by the utilization of adaptive strategies in data augmentation, we hypothesize that batch complexity could also have a positive influence on model regularization. Thus, we add a batch complexity scalar weight to the intermediate CTC loss regularization method. Then, we combine the two methods to form the proposed complexity-based adaptive training for better ASR performance. 

\subsection{Sample MinMax-IBF policy}

\begin{table}[h]\footnotesize
\vspace{-0.3cm}
\centering
\renewcommand{\arraystretch}{1.5}
\caption{Comparison of the augmentation tensity between SapAugment and the proposed MinMax-IBF.}
\setlength{\tabcolsep}{2mm}{
    \begin{tabular}{c | c | c}
    \hline
    Batch loss values & Loss Rank & MinMax IBF\\ 
    \hline
    (1,2,6) & $(\frac{1}{3}, \frac{2}{3}, \frac{3}{3})$ & $(0, \frac{1}{5}, \frac{5}{5})$ \\
    (1,5,6) & $(\frac{1}{3}, \frac{2}{3}, \frac{3}{3})$ & $(0, \frac{4}{5}, \frac{5}{5})$ \\
    \hline
    \end{tabular}
}
\vspace{-0.2cm}
\end{table}

As mentioned in equation (1), SapAugment employs the ranking of loss (\(\mathbf{l}_{\text{rank}}\) = 1,2,...,\(\mathbf{B}\)) in the mini-batch for the incomplete beta function. More precisely, the relative sample complexity (\(\mathbf{l}_{\text{rank}}\)/ \(\mathbf{B}\)) decides the amount of those augmentation methods.
In table 2, consider two batches, each comprising three samples. The loss values for the samples are (1, 2, 6) and (1, 5, 6) respectively. Despite having the same loss rank, the second sample in each batch exhibits a significant difference in loss values. Ideally, the sample with a loss value of 2 should receive a higher amount of augmentation, while the sample with a loss value of 5 should receive less augmentation. 

We propose MinMax-IBF policy giving a new normalization formula for each \(\mathbf{x}_{\text{i}}\) in the incomplete beta function. It not only calculates the complexity of each sample, as shown in Table 2, but can also reflect in detail the influence of the relative magnitude of the sample loss on the complexity. This makes the complexity estimation more generalized and robust.

\begin{equation}
    x_{\text{i}} =  \frac{L_{\text{i}} - L_{\text{min}} }{L_{\text{max}} - L_{\text{min}}},\ x_{\text{i}} \in [0,1]
\end{equation}
where \(\mathbf{L}_{\text{i}}\) is the loss value of the \(\mathbf{i}_{\text{th}}\) sample in the min-batch, \(\mathbf{L}_{\text{min}}\) and \(\mathbf{L}_{\text{max}}\) represent the minimum and maximum loss value within the mini-batch, respectively.

The adaptive factor of data augmentation (DA) is calculated as follows. Each training sample is applied data augmentation of an appropriate intensity according to the adaptive factor.
\begin{equation}
    f_{\text{DA}} = 1 - IBF(x_{\text{i}}),\ f_{\text{DA}} \in [0,1]
\end{equation}

With the proposed formulas (6) and (7), we get the sample policy $f_{\text{DA}}$ and perform a linear mapping to determine the number of data augmentation iterations. For example, the number of time masks can be calculated as 4 times the value of $f_{\text{DA}}$. we still ensure that difficult samples receive fewer augmentations while easy samples receive more augmentations.

\subsection{Batch Complexity Regularization}
In equation 5, when using the intermediate loss regularization, the total loss of the multi-layer network is the weighted sum of the sub-model and total model loss, where the weight \(\lambda\) is a hyper-parameter balancing between loss from the last encoder layer and intermediate layers. 

Additionally, we use a scalar weight to include the batch complexity information. When the batch complexity decreased, intermediate regularization would increase. For easy samples, it becomes crucial to emphasize the learning of the sub-model, specifically the early layer within the model. Our objective is for the early layer to be capable of recognizing the samples even before they reach the upper layers. Conversely, for difficult samples, the adaptive scalar weight ensures that the total loss remains within a suitable range.

The adaptive factor of InterCTC is calculated as follows, where B represents the batch size.
\begin{equation}
    f_{\text{CTC}} = \frac{\sum_{i=1}^{B} (1- IBF(x_{\text{i}})) }{B}
\end{equation}

The proposed loss of the network is:

\begin{equation}
    L_{\text{total}} = (1- \lambda)L_{\text{CTC}} + f_{\text{CTC}} * \lambda {L_{\text{InterCTC}}}
\end{equation}

Note that, the \(\mathbf{L}_{\text{InterCTC}}\) could be from multiple sub-models of the network, more specifically, the average of those sub-models. Suppose the number of sub-models is N and the loss from sub-model n is \(\mathbf{L}_{n}\), we compute the InterCTC loss as:

\begin{equation}
  L_{\text{InterCTC}} =  \frac{1}{N} \sum_{n=1}^{N} L_{\text{n}} (y|x_{\text{n}})
\end{equation}

\subsection{The CBA Training Method}

We design a two-stage continued training process. In the first stage, we do not use MinMax-IBF policy. Instead, we employ the intermediate CTC regularization. It helps the model fit and enhance the bottom encoder layers, in the early training time. Moreover, as the training process goes on, the regularization nature also eases the overfitting problem. SpecAugment is applied in this stage and also the next stage. Note that, the first training stage is not adaptive.

In the second stage, we continue to train the model with the MinMax-IBF policy for both data augmentation and intermediate loss regularization. By adding the adaptive scalar weight to the original intermediate CTC loss, the regularization density now depends on the complexity of each batch. After continuing training for certain epochs, we then average 10 models selected by cross-validation loss. The final model is ready to use.

Figure 2 demonstrates the details of the second adaptive training stage. Specifically, it consists of two passes. In addition to the regular forward and backward computations, it requires an extra forward pass. This forward pass calculates the loss, which is used to guide data augmentation and regularization tensity.

Note that, we do not use MinMax-IBF policy in the first training stage. The reason behind this is that, during the early stages of model training, as the model is still far from convergence, there tend to be significant fluctuations in the loss values generated by the samples. If our MinMax-IBF policy is employed in the batch, it may lead to the following issues: 1. Within a batch, the distribution of augmentation tensity tends to be biased towards extreme values. 2. Across different batches, due to the large fluctuations in loss values, a sample that is considered difficult in one batch may be classified as easy in another batch, which is not reasonable.

\begin{figure}
\centering
\includegraphics[width=7cm]{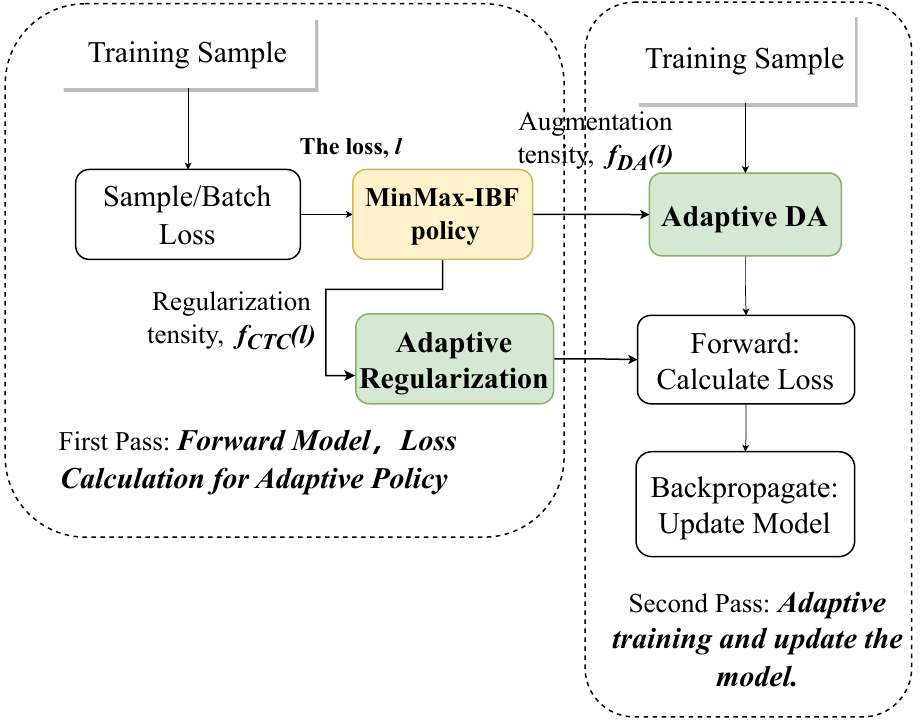}
\caption{The two passes for the continued training stage.}
\label{fig:label}
\end{figure}

\section{Experiments \& Analysis}

\subsection{Dataset}
We perform our experiments on two datasets: AISHELL-1 \cite{bu2017aishell} and LibriSpeech \cite{panayotov2015librispeech}. AISHELL-1 is an open-source Mandarin speech corpus containing 120,098 utterances from 340 speakers in the training set; the development set contains 14,326 utterances from 40 speakers; the test set contains 7,176 utterances from 20 speakers. LibriSpeech contains approximately 1,000 hours of speech from public audiobooks. We use the US accent 100-hour subset of LibriSpeech for training efficiency and report our results on the test-clean and test-other sets.

\subsection{Implementation}
We chose the WeNet toolkit \cite{yao2021wenet} to implement our experiments. For the acoustic model, we utilize Conformer in WeNet toolkit. Following the network architecture configuration (train\_conformer), the acoustic feature is 80-dim FBANK computed with a 25ms window and a 10ms shift. We conduct experiments with a 12-layer Conformer encoder, a 6-layer Transformer decoder, and the attention-rescore decoding method, without any additional language model. We average the model at the end of the training stage according to the cross-validation loss. For LibriSpeech, the model is trained for 120 epochs, and for AISHELL-1, the model is trained for 240 epochs. We use the SpecAugment methods in the WeNet directly. We also implement the SapAugment policy based on the WeNet Conformer. In terms of hyperparameter setting, IBF's parameters are consistent with SapAugment \cite{hu2021sapaugment}, with s set to 0.5, a set to 5, and the subModel selecting layers 6, 9 and 12 from the 12-layer encoder.

\begin{table}[h]\footnotesize
\vspace{-0cm}
\centering
\caption{CER performance of the proposed method against baselines trained and tested on LibirSpeech (100h) and AISHELL-1 dataset. The numbers in the table refer to the percentage of Word Error Rate (LibriSpeech) or Character Error Rate (AISHELL-1).}
\renewcommand{\arraystretch}{1.3}
\fontsize{9}{9}\selectfont
\setlength{\tabcolsep}{1mm}{
    \begin{tabular}{lcc|c}
    \toprule
    \multicolumn{1}{l}{\textbf{Method}} & \multicolumn{2}{c}{\textbf{LibriSpeech 100h}} & \multicolumn{1}{c}{\textbf{AISHELL-1}} \\ 
    \cmidrule(l){2-4} 
    WER/CER & test-clean  & test-other & test \\
    \hline 
    SpecAugment  & 8.36    & 23.34  & 4.62  \\ 
    SapAugment \cite{hu2021sapaugment}  & 8.30  & 21.50  & 4.60   \\ 
    Intermediate CTC   &  7.89   & 22.47  & 4.54 \\
    \textbf{CBA Training}   & \textbf{7.24}   & \textbf{20.04}   & \textbf{4.33}  \\ \hline
    \end{tabular}
}
\vspace{-0.2cm}
\end{table}

\subsection{Evaluation}

We conduct all experiments on the WeNet Conformer, which implements the SpecAugment method. As shown in Table 3, Row 1 is with the SpecAugment, which is our augmentation baseline, without any policy for data augmentation. Row 2 is the performance of SapAugment from \cite{hu2021sapaugment}. Row 3 only uses the intermediate CTC loss regularization in \cite{lee2021intermediate}, without the adaptive scalar weight. The last row shows the performance of our CBA training method. The proposed CBA training method demonstrates a relative 13.4\% and 14.1\% improvement in WER compared to the baseline SpecAugment of WeNet Conformer, achieving WER 7.24\% and 20.04\% in the LibriSpeech test-clean and test-other datasets, respectively. 

\begin{table}[h]\footnotesize
\vspace{-0.2cm}
\centering
\caption{Words Error Rate (WER) in ablation study on LibirSpeech test-clean and test-other sets.}
\renewcommand{\arraystretch}{1.3}
\fontsize{9}{9}\selectfont
\setlength{\tabcolsep}{1.5mm}{
    \begin{tabular}{l cc}
        \toprule
        \multirow{2}{*}{\textbf{Method/Dataset}} & \multicolumn{2}{c}{\textbf{LibriSpeech 100h}} \\
        \cmidrule(l){2-3}
        & test-clean  & test-other \\ \hline
        SpecAug, without policy   &8.36 &23.34 \\ 
        \hspace{0.5cm} + MinMax-IBF DA  &7.87 &22.55 \\ 
        \hspace{1cm} + Regularization(R)  &7.80  &22.23 \\
        \hspace{1.5cm} + \textbf{Adaptive R (AR)}  &\textbf{7.24}  &\textbf{20.04} \\
        \hline
    \end{tabular}
}
\vspace{-0.2cm}
\end{table}

\begin{table}[h]\footnotesize
\vspace{-0.2cm}
\centering
\caption{Character Error Rate (CER) in ablation study on AISHELL-1 dataset. The experiments are carried out with the same method as LibriSpeech.}
\renewcommand{\arraystretch}{1.3}
\fontsize{9}{9}\selectfont
\setlength{\tabcolsep}{1.5mm}{
    \begin{tabular}{lc}
        \toprule
        \multirow{1}{*}{\textbf{Method/Dataset}}  & \multicolumn{1}{c}{\textbf{AISHELL-1 test}} \\
        \hline
        SpecAug, without policy   &  4.62    \\
        \hspace{0.5cm} + MinMax-IBF DA  & 4.43 \\ 
        \hspace{1cm} + Regularization(R)  & 4.42   \\
        \hspace{1.5cm} + \textbf{Adaptive R (AR)}  & \textbf{4.33} \\
        \hline
    \end{tabular}
}
\vspace{-0.5cm}
\end{table}

\subsection{Ablation Study}

In this section, we conduct ablation studies on the LibriSpeech 100h and AISHELL-1 dataset to examine the individual contributions of each component of the proposed complexity-boosted adaptive training.
In Table 4 and Table 5, Row 1 uses the SpecAugment only. In Row 2, the MinMax-IBF policy is added to SpecAugment for data augmentation (DA) in the continued training stage. In Row 3, SpecAugment incorporates intermediate CTC loss regularization (R). It is worth mentioning that this regularization is not yet adaptive. Based on Row 3, Row 4 uses our complexity-boosted adaptive regularization method (AR) in the continued training stage. It trains with the intermediate CTC regularization method first and then continues training with the MinMax-IBF policy for DA and regularization, which is the proposed CBA training. Further observations are: 

\begin{itemize}

\item Augmentation with the proposed MinMax-IBF policy on top of SpecAugment further improves performance, with the proposed two-stage continued training method. This shows consistency in LibriSpeech 100h and AISHELL-1 test set. The adaptation policy gives proper augmentation according to data complexity. This helps model convergence after learning basic data distribution.   

\item Utilizing the proposed complexity regularization method without the adaptive data augmentation also performs better on two datasets than the SpecAugment baseline. The adaptive loss helps regularize the model well in the whole training process.

\item  By combining the two methods with the proposed continued training, we improve performance largely, which is better than either of the proposed methods. In the first training stage, the model benefits from the original data distribution without adaptation. In the second stage, the proposed adaptive regularization method reduces overfitting.
\end{itemize}

\begin{table}[h]\footnotesize
\vspace{-0.2cm}
\centering
\caption{Performance using the proposed two-stage training method vs. training from scratch. }
\renewcommand{\arraystretch}{1.5}
\fontsize{9}{10}\selectfont
\setlength{\tabcolsep}{1mm}{
    \begin{tabular}{lcc|c}
        \hline
        \multirow{2}{*}{\textbf{Method/Dataset}} & \multicolumn{2}{c}{\textbf{LibriSpeech 100h}}  & \multicolumn{1}{c}{\textbf{AISHELL-1}} \\
        \cmidrule(l){2-4}
         & clean  & other & test \\ \hline
        FS $\cup$ MinMax-IBF DA  & 8.26  & 22.94 & 4.57 \\
        CT $\cup$ MinMax-IBF DA  & \textbf{7.87} & \textbf{22.55} & \textbf{4.43} \\ 
        \hline
        FS $\cup$ MinMax-IBF DA+AR & 7.72  & 22.2 & 4.47 \\ 
        CT $\cup$ MinMax-IBF DA+AR & \textbf{7.24}  & \textbf{20.04}  & \textbf{4.33} \\ \hline
    \end{tabular}
}
\vspace{-0.2cm}
\end{table}

Table 6 presents the efficacy of the two-stage training approach. For the purpose of comparison, it is important to note that each row in this table corresponds to an equal number of total training epochs. Row 1 uses MinMax-IBF policy for data augmentation (DA) and trains from scratch (FS). Row 2 uses the policy only in the continued training (CT) stage. Row 3 uses the policy for both DA and adaptive intermediate loss regularization (AR) from scratch. Row 4 is the CBA training method. Whether utilizing Aug with MinMax-IBF policy alone or both MinMax-IBF policy and adaptive regularization, the two-stage training method yields superior results on the LibriSpeech 100h and AISHELL-1 datasets.

Note that, in this table, intermediate CTC loss regularization is employed during the whole training process, including the continued training stage. We add the adaptive weight from the second training stage. This relieves the overfitting problem in the training process, and also adaptively motivates the information from early layers flowing to the upper layers. When continuing the training, we begin to use the adaptive policy. If it is applied in the early stage, there will be excessive fluctuations in the loss of the data. The values of these losses may not fall within a reasonable range of policy inputs, like mentioned in section 3.3.
Moreover, it reduces the diversity of the data, leading to a convergence towards a singular data distribution, which hinders the learning process.

\section{CONCLUSION}

This paper focuses on a complexity-boosted adaptive training method for better ASR performance in low-resource scenarios. Initially, we apply regularization to the training process by utilizing the intermediate CTC loss regularization. Subsequently, we proceed with training using the proposed MinMax-IBF policy for adaptive data augmentation and regularization. This continued training method has proved effective on both Mandarin and English datasets, AISHELL-1 and LibriSpeech, without any parameter increase or computation burden when inference. Our method exceeds WeNet Conformer SpecAugment baseline by a relative 13.4\% and 14.1\% WER drop on LibriSpeech(100h) and a relative 6.3\% CER improvement on AISHELL-1. Future work will investigate the effects of applying adaptive training within different internal modules. 


\begin{thebibliography}{10}

\bibitem{chan2016listen}
William Chan, Navdeep Jaitly, Quoc Le, and Oriol Vinyals,
\newblock ``Listen, attend and spell: A neural network for large vocabulary conversational speech recognition,''
\newblock in {\em 2016 IEEE international conference on acoustics, speech and signal processing (ICASSP)}. IEEE, 2016, pp. 4960--4964.

\bibitem{chorowski2015attention}
Jan~K Chorowski, Dzmitry Bahdanau, Dmitriy Serdyuk, Kyunghyun Cho, and Yoshua Bengio,
\newblock ``Attention-based models for speech recognition,''
\newblock {\em Advances in neural information processing systems}, vol. 28, 2015.

\bibitem{bartelds2023making}
Martijn Bartelds, Nay San, Bradley McDonnell, Dan Jurafsky, and Martijn Wieling,
\newblock ``Making more of little data: Improving low-resource automatic speech recognition using data augmentation,''
\newblock {\em arXiv preprint arXiv:2305.10951}, 2023.

\bibitem{casanova2022asr}
Edresson Casanova, Christopher Shulby, Alexander Korolev, Arnaldo~Candido Junior, Anderson da~Silva Soares, Sandra Alu{\'\i}sio, and Moacir~Antonelli Ponti,
\newblock ``Asr data augmentation in low-resource settings using cross-lingual multi-speaker tts and cross-lingual voice conversion,''
\newblock {\em arXiv preprint arXiv:2204.00618}, 2022.

\bibitem{lam2021fly}
Tsz~Kin Lam, Mayumi Ohta, Shigehiko Schamoni, and Stefan Riezler,
\newblock ``On-the-fly aligned data augmentation for sequence-to-sequence asr,''
\newblock {\em arXiv preprint arXiv:2104.01393}, 2021.

\bibitem{hu2021sapaugment}
Ting-Yao Hu, Ashish Shrivastava, Jen-Hao~Rick Chang, Hema Koppula, Stefan Braun, Kyuyeon Hwang, Ozlem Kalinli, and Oncel Tuzel,
\newblock ``Sapaugment: Learning a sample adaptive policy for data augmentation,''
\newblock in {\em ICASSP 2021-2021 IEEE International Conference on Acoustics, Speech and Signal Processing (ICASSP)}. IEEE, 2021, pp. 4040--4044.

\bibitem{park2019specaugment}
Daniel~S Park, William Chan, Yu~Zhang, Chung-Cheng Chiu, Barret Zoph, Ekin~D Cubuk, and Quoc~V Le,
\newblock ``Specaugment: A simple data augmentation method for automatic speech recognition,''
\newblock {\em arXiv preprint arXiv:1904.08779}, 2019.

\bibitem{kim2022squeezeformer}
Sehoon Kim, Amir Gholami, Albert Shaw, Nicholas Lee, Karttikeya Mangalam, Jitendra Malik, Michael~W Mahoney, and Kurt Keutzer,
\newblock ``Squeezeformer: An efficient transformer for automatic speech recognition,''
\newblock {\em Advances in Neural Information Processing Systems}, vol. 35, pp. 9361--9373, 2022.

\bibitem{kim2023branchformer}
Kwangyoun Kim, Felix Wu, Yifan Peng, Jing Pan, Prashant Sridhar, Kyu~J Han, and Shinji Watanabe,
\newblock ``E-branchformer: Branchformer with enhanced merging for speech recognition,''
\newblock in {\em 2022 IEEE Spoken Language Technology Workshop (SLT)}. IEEE, 2023, pp. 84--91.

\bibitem{ren2022improving}
Xiaoming Ren, Huifeng Zhu, Liuwei Wei, Minghui Wu, and Jie Hao,
\newblock ``Improving mandarin speech recogntion with block-augmented transformer,''
\newblock {\em arXiv preprint arXiv:2207.11697}, 2022.

\bibitem{burchi2021efficient}
Maxime Burchi and Valentin Vielzeuf,
\newblock ``Efficient conformer: Progressive downsampling and grouped attention for automatic speech recognition,''
\newblock in {\em 2021 IEEE Automatic Speech Recognition and Understanding Workshop (ASRU)}. IEEE, 2021, pp. 8--15.

\bibitem{ko2015audio}
Tom Ko, Vijayaditya Peddinti, Daniel Povey, and Sanjeev Khudanpur,
\newblock ``Audio augmentation for speech recognition,''
\newblock in {\em Sixteenth annual conference of the international speech communication association}, 2015.

\bibitem{6639100}
Michael~L. Seltzer, Dong Yu, and Yongqiang Wang,
\newblock ``An investigation of deep neural networks for noise robust speech recognition,''
\newblock in {\em 2013 IEEE International Conference on Acoustics, Speech and Signal Processing}, 2013, pp. 7398--7402.

\bibitem{7953152}
Tom Ko, Vijayaditya Peddinti, Daniel Povey, Michael~L. Seltzer, and Sanjeev Khudanpur,
\newblock ``A study on data augmentation of reverberant speech for robust speech recognition,''
\newblock in {\em 2017 IEEE International Conference on Acoustics, Speech and Signal Processing (ICASSP)}, 2017, pp. 5220--5224.

\bibitem{graves2006connectionist}
Alex Graves, Santiago Fern{\'a}ndez, Faustino Gomez, and J{\"u}rgen Schmidhuber,
\newblock ``Connectionist temporal classification: labelling unsegmented sequence data with recurrent neural networks,''
\newblock in {\em Proceedings of the 23rd international conference on Machine learning}, 2006, pp. 369--376.

\bibitem{graves2012sequence}
Alex Graves,
\newblock ``Sequence transduction with recurrent neural networks,''
\newblock {\em arXiv preprint arXiv:1211.3711}, 2012.

\bibitem{vaswani2017attention}
Ashish Vaswani, Noam Shazeer, Niki Parmar, Jakob Uszkoreit, Llion Jones, Aidan~N Gomez, {\L}ukasz Kaiser, and Illia Polosukhin,
\newblock ``Attention is all you need,''
\newblock {\em Advances in neural information processing systems}, vol. 30, 2017.

\bibitem{toshniwal2017multitask}
Shubham Toshniwal, Hao Tang, Liang Lu, and Karen Livescu,
\newblock ``Multitask learning with low-level auxiliary tasks for encoder-decoder based speech recognition,''
\newblock {\em arXiv preprint arXiv:1704.01631}, 2017.

\bibitem{lee2021intermediate}
Jaesong Lee and Shinji Watanabe,
\newblock ``Intermediate loss regularization for ctc-based speech recognition,''
\newblock in {\em ICASSP 2021-2021 IEEE International Conference on Acoustics, Speech and Signal Processing (ICASSP)}. IEEE, 2021, pp. 6224--6228.

\bibitem{gulati2020conformer}
Anmol Gulati, James Qin, Chung-Cheng Chiu, Niki Parmar, Yu~Zhang, Jiahui Yu, Wei Han, Shibo Wang, Zhengdong Zhang, Yonghui Wu, et~al.,
\newblock ``Conformer: Convolution-augmented transformer for speech recognition,''
\newblock {\em arXiv preprint arXiv:2005.08100}, 2020.

\bibitem{yao2021wenet}
Zhuoyuan Yao, Di~Wu, Xiong Wang, Binbin Zhang, Fan Yu, Chao Yang, Zhendong Peng, Xiaoyu Chen, Lei Xie, and Xin Lei,
\newblock ``Wenet: Production oriented streaming and non-streaming end-to-end speech recognition toolkit,''
\newblock {\em arXiv preprint arXiv:2102.01547}, 2021.

\bibitem{incompletBeta}
R.B. Paris,
\newblock {\em Incomplete beta functions},
\newblock NIST Handbook of Mathematical Functions, Cambridge University Press, 2010.

\bibitem{bu2017aishell}
Hui Bu, Jiayu Du, Xingyu Na, Bengu Wu, and Hao Zheng,
\newblock ``Aishell-1: An open-source mandarin speech corpus and a speech recognition baseline,''
\newblock in {\em 2017 20th conference of the oriental chapter of the international coordinating committee on speech databases and speech I/O systems and assessment (O-COCOSDA)}. IEEE, 2017, pp. 1--5.

\bibitem{panayotov2015librispeech}
Vassil Panayotov, Guoguo Chen, Daniel Povey, and Sanjeev Khudanpur,
\newblock ``Librispeech: an asr corpus based on public domain audio books,''
\newblock in {\em 2015 IEEE international conference on acoustics, speech and signal processing (ICASSP)}. IEEE, 2015, pp. 5206--5210.

\end{thebibliography}

\end{document}